\documentclass{Interspeech}

\interspeechcameraready

\title{IndexTTS: An Industrial-Level Controllable and Efficient Zero-Shot Text-To-Speech System}

\author{Wei}{Deng}
\author{Siyi}{Zhou}
\author{Jingchen}{Shu}
\author{Jinchao}{Wang}
\author{Lu}{Wang}

\affiliation{Artificial Intelligence Platform Department}{bilibili}{China}
\email{\{xuanwu,zhousiyi02,shujingchen,wangjinchao,wanglu08\}@bilibili.com}

\usepackage{comment}
\usepackage{graphicx}
\usepackage{booktabs}
\usepackage{multirow}
\usepackage{CJKutf8}
\begin{document}

\maketitle

\begin{abstract}
    
    Recently, large language model (LLM) based text-to-speech (TTS) systems have gradually become the mainstream in the industry due to their high naturalness and powerful zero-shot voice cloning capabilities.
    
    Here, we introduce the IndexTTS system, which is mainly based on the XTTS and Tortoise model. 
    We add some novel improvements. Specifically, in Chinese scenarios, we adopt a hybrid modeling method that combines characters and pinyin, 
    making the pronunciations of polyphonic characters and long-tail characters controllable. 
    We also performed a comparative analysis of the Vector Quantization (VQ) with Finite-Scalar Quantization (FSQ) for codebook utilization of acoustic speech tokens. To further enhance the effect and stability of voice cloning, we introduce a conformer-based speech conditional encoder and replace the speechcode decoder with BigVGAN2.
    
    Compared with XTTS, it has achieved significant improvements in naturalness, content consistency, and zero-shot voice cloning. As for the popular TTS systems in the open-source, such as Fish-Speech, CosyVoice2, FireRedTTS and F5-TTS, IndexTTS has a relatively simple training process, more controllable usage, and faster inference speed.  Moreover, its performance surpasses that of these systems. Our demos are available at \url{https://index-tts.github.io}.
\end{abstract}

\section{Introduction}

Text-to-speech synthesis (TTS) has extensive applications in fields such as human-computer interaction, education, and entertainment.
For example, in video creation scenarios in recent years, TTS can assist users in quickly generating video dubbing, saving recording time, and thus playing a crucial role in the creation process.
Many creators hope to provide personalized and highly natural speech synthesis services to meet the needs of different scenarios. 

The TTS system, which is based on large language models and can be trained using massive amounts of general speech data, demonstrates impressive performance in speech generation, such as XTTS\cite{casanova2024xtts}, Fish-Speech\cite{liao2024fish}, CosyVoice2\cite{du2024CosyVoice}, FireRedTTS\cite{guo2024fireredtts} and F5-TTS\cite{chen2024f5}. Compared to traditional systems that rely on more intricate manual designs, such as Mega-tts 2\cite{jiang2023mega} and Yourtts\cite{casanova2022yourtts}, these systems have achieved significant improvements in naturalness, particularly in zero-shot voice cloning. Generative TTS powered by big data can be roughly classified into three categories. The first is the neural codec language model. To ensure the quality of synthesized audio, it typically employs a multi-codebook codec along with a high-frame-rate configuration, like in Vall-E\cite{wang2023neural}. This architecture is simple and straightforward, yet it has drawbacks: longer training and inference times, along with compromised stability. The second is end-to-end diffusion-based TTS, a non-autoregressive (NAR) model. F5-TTS\cite{chen2024f5} and Seed-TTS\cite{anastassiou2024seed} are case-in-point. This approach yields high-quality synthesized audio and is suitable for voice editing but is difficult to stream, so not for real-time use. Finally, the hybrid architecture typically uses a single codebook and a low bitrate codec, generating high-quality audio through a standalone decoder such as diffusion or HiFiGAN\cite{kong2020hifi}. It balances performance and generation quality and offers good stability. Due to the success of large language models, tokenization is the trend of the future. For industrial-level applications, stability is crucial. Here, we opt for the hybrid architecture, using a single codebook codec and reconstruct high quality voice through a speech decoder, such as XTTS, Fish-Speech and CosyVoice2. 

Based on XTTS\cite{casanova2024xtts} and Tortoise\cite{betker2023better}, we have made several improvements, which mainly include the following:
We remove the front-end G2P module and use raw text as input, along with a BPE-based text tokenizer. This simplifies input preprocessing, facilitates multi-language expansion, and enables end-to-end learning of word or polyphone pronunciations via big data context integration. To address the pronunciation control of polyphones and low-frequency characters in Chinese scenarios, which inevitably occur in real world video creation, we propose a hybrid character-pinyin modeling approach. This allows video creators to correct pronunciations by directly inputting pinyin. Moreover, VQ\cite{van2017neural} may suffer from low-utilization of the quantization codebook due to codebook collapse. we conducted a comparative analysis between VQ and FSQ\cite{mentzer2023finite} in terms of their codebook utilization for acoustic token representation, achieving nearly 100\% codebook utilization. Finally, we have made significant improvements in prosody naturalness, the similarity of zero-shot voice cloning, and system stability. The main improvements and contributions are summarized as follows:
\begin{itemize}
    \item In Chinese scenarios, we have introduced a character-pinyin hybrid modeling approach. This allows for quick correction of mispronounced characters.
    \item We develop the IndexTTS system, incorporating a conformer conditioning encoder and a BigVGAN2\cite{lee2022bigvgan}-based speechcode decoder. This improves training stability, voice timbre similarity, and sound quality.
    \item We release all test sets, including those for polysyllabic words, subjective and objective test sets\footnote{\url{https://github.com/index-tts/index-tts}}.
\end{itemize}

\section{IndexTTS System}
\begin{figure*}[t]
  \centering
  \includegraphics[scale=0.55]{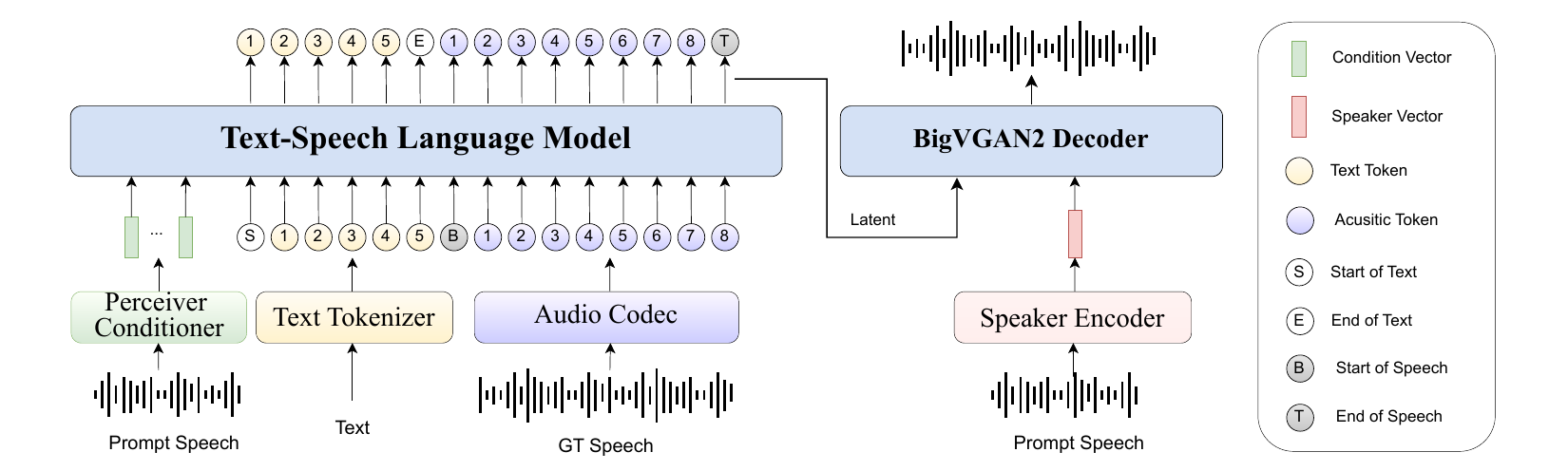}
  \caption{An overview of IndexTTS, a text-to-speech language model conditioned on prompt speech and text tokens generates acoustic tokens, and the BigVGAN2 decoder convert the LLM output latent into waveform.}
  \label{fig:indextts}
\end{figure*}

Similar to XTTS\cite{casanova2024xtts}, our system incorporates speech-to-codec VQVAE\cite{van2017neural} codec, text-to-codec language model and latent-to-audio decoder, as depicted in Figure~\ref{fig:indextts}.

\begin{table*}[!htbp]
\caption{Preprocessing Examples for Training Samples Combining Chinese Characters and Pinyin}
  \label{tab:text_tokenizer}
  \centering
\begin{tabular}{@{}lc@{}}
\toprule
\textbf{Input: } & \begin{CJK*}{UTF8}{gbsn}晕眩是一种感觉，\end{CJK*}I want to go to the supermarket!                                                              \\ \midrule
\textbf{Mix Pinyin: }   &  \begin{CJK*}{UTF8}{gbsn}晕\ XUAN4\ 是\ 一\ 种\ GAN3\ 觉\ ，\end{CJK*}\ I WANT TO GO TO THE SUPERMARKET !                                                      \\ \midrule
\textbf{BPE Tokens: } &  \begin{CJK*}{UTF8}{gbsn}\_晕,\ \_XUAN4,\ \_是,\ \_一,\ \_种,\ \_GAN3,\ \_觉,\ \_，\end{CJK*}, \_I, \_WANT, \_TO, \_GO, \_TO, \_THE, \_SUPER, M, AR, KE, T, ! \\ \bottomrule
\end{tabular}
\end{table*}

\subsection{Text tokenizer}
Currently, our system only supports two languages, Chinese and English. We directly use the raw text as input, which is tokenized by a BPE-based text tokenizer, This makes it convenient to extend the system to other languages. Due to the large number of polyphonic characters in Chinese, we adopt a hybrid-modeling approach of Chinese characters and pinyin in Chinese-related scenarios. The vocabulary size of the text tokenizer is 12,000. It encompasses 8,400 Chinese characters along with their corresponding 1,721 pinyin, English word pieces, and several special symbols. During specific training, we randomly select a certain proportion of non-polyphonic characters and replace them with pinyin. An example of preprocessing process is presented in Table~\ref{tab:text_tokenizer}.

\subsection{Neural Speech Tokenizer}
Vector Quantization (VQ) is a powerful tool for speech coding, but it may suffer from codebook collapse\cite{mentzer2023finite}, The codebook utilization of VQ and FSQ was analyzed in the following experiments. We increased the parameters of the Variational Autoencoder (VAE) to around 50M. The VAE receives a mel-spectrogram as input and encodes each frame with VQ using approximately 8192 codes. The sampling rate of the input audio is 24 kHz, and the token rate output by the speech tokenizer is 25 Hz.

\begin{table}[!htbp]
\caption{Error and Correction Statistics for Polyphonic Character Pronunciation}
  \label{tab:polyphonic}
  \centering
\begin{tabular}{@{}lcc@{}}
\toprule
               & \textbf{Sentences} & \textbf{Percentage} \\ \midrule
\textbf{Total} & 2500               & 100\%               \\
\textbf{A1}    & 465                & 18.6\%              \\
\textbf{A2}    & 437                & 94.0\%               \\ \bottomrule
\end{tabular}
\end{table}

\begin{table*}[h]
  \caption{Word Error Rate (WER) and Speaker Similarity (SS) Results for IndexTTS and Baseline Models}
  \label{tab:wer_ss_results}
  \centering
  \begin{tabular}{@{}lcccccccccc@{}}
\toprule
\multirow{2}{*}{\textbf{Model}} & \multicolumn{2}{c}{\textbf{aishell1\_test}} & \multicolumn{2}{c}{\textbf{commonvoice\_zh}} & \multicolumn{2}{c}{\textbf{commonvoice\_en}} & \multicolumn{2}{c}{\textbf{librispeech\_test\_clean}} & \multicolumn{2}{c}{\textbf{AVG}} \\ \cmidrule(lr){2-3} \cmidrule(lr){4-5} \cmidrule(lr){6-7} \cmidrule(lr){8-9} \cmidrule(lr){10-11}
                       & \textbf{CER(\%)↓ }         & \textbf{SS↑}           & \textbf{CER(\%)↓}                & \textbf{SS↑}                 & \textbf{WER(\%)↓}                & \textbf{SS↑}                 & \textbf{WER(\%)↓}               & \textbf{SS↑}                 & \textbf{WER(\%)↓}     & \textbf{SS↑}      \\ \cmidrule(){1-11}
\textbf{Human}                  & 2.0               & 0.846          & 9.5       & 0.809       & 10.0             & 0.820              & 2.4           & 0.858               & 5.1          & 0.836   \\
\cmidrule(){1-11}
\textbf{CosyVoice2}            & 1.8               & \textbf{0.796}          & 9.1                     & 0.743               & 7.3                     & 0.742               & 4.9                    & \textbf{0.837}               & 5.9          & \textbf{0.788}    \\
\textbf{F5TTS}                  & 3.9               & 0.743          & 11.7                    & \textbf{0.747}               & 5.4                     & 0.746               & 7.8                    & 0.828               & 8.2          & 0.779    \\
\textbf{Fishspeech}             & 2.4               & 0.488          & 11.4                    & 0.552               & 8.8                     & 0.622               & 8.0                    & 0.701               & 8.3          & 0.612    \\
\textbf{FireRedTTS}             & 2.2               & 0.579          & 11.0                    & 0.593               & 16.3                    & 0.587               & 5.7                    & 0.698               & 7.7          & 0.631    \\
\textbf{XTTS}                   & 3.0               & 0.573          & 11.4                    & 0.586               & 7.1                     & 0.648               & 3.5                    & 0.761               & 6.0          & 0.663    \\ \cmidrule(){1-11}
\textbf{IndexTTS}          & \textbf{1.3}               & 0.744          & \textbf{7.0}                     & 0.742               & \textbf{5.3}                     & \textbf{0.758}               & \textbf{2.1}                    & 0.823               & \textbf{3.7}          & 0.776    \\ \bottomrule
\end{tabular}
\end{table*}

\subsection{Large Language Model for TTS}
The text-to-codec large language model (LLM) is based on the decoder-only transformer architecture, similar to XTTS.
It generates a series of audio mel tokens from the input series of text tokens.
The LLM is also conditioned by a transformer-based conditioning encoder, which we replace with a Conformer encoder with a subsample rate of 2.
We found that this replacement can enhance timbre similarity and training stability.

The training processes of conditional LLM can be broadly categorized into the following types, The input sequence is structured as follows (\textbf{[BT]} \textbf{[ET]} indicate the beginning and end of the text token sequence. \textbf{[BA]} and \textbf{[EA]} denote the start and end of the audio token sequence):
\begin{itemize}
    \item \textbf{SEQ1:} ``\textbf{[BT], prompt\_text, text, [ET], [BA], prompt\_audio, audio, [EA]}'', such as Vall-E and Fish-Speech, it concatenates all the tokens of the prompt and the target.
    \item \textbf{SEQ2:} ``\textbf{[BT], text, [ET], [BA], audio, [EA]}'', for instance, CosyVoice2 directly generates audio tokens from the text tokens series.
    \item \textbf{SEQ3:} ``\textbf{speaker\_info, [BT], text, [ET], [BA], audio, [EA]}'', for example, in XTTS\cite{casanova2024xtts}, CosyVoice\cite{du2024CosyVoice} and Tortoise\cite{betker2023better}, the speaker information of the prompt audio is compressed into one or 32 latent vectors, which serve as the conditions for the LLM.
\end{itemize}

\textbf{SEQ1} and \textbf{SEQ2} must rely on the text corresponding to the prompt audio during the inference process. The inference input prefix sequence can be constructed as ``\textbf{[BT], prompt\_text, text, [ET], [BA], prompt\_audio}''. In comparison, \textbf{SEQ3} only requires the prompt audio. The inference input prefix sequence is ``\textbf{speaker\_info, [BT], text, [ET], [BA]}''. The autoregressive generation of LM is started from such input prefix sequence until the ``End of sequence'' token ``\textbf{[EA]}'' is detected. 

We adopt the \textbf{SEQ3}. It is worth emphasizing that not relying on prompt text is crucial in certain scenarios. For example, in cross-language voice cloning, if prompt text must be provided or identified through a multilingual ASR system, its usability will be significantly limited. Additionally, conditioning on both the prompt text and the audio token series will substantially increase the inference time.

We also found that, compared to single-speaker encoding vectors such as Tortoise \cite{betker2023better} and CosyVoice \cite{du2024CosyVoice}, or speech-prompting methods like Vall-E, the Conformer-based Perceiver demonstrates superior ability in capturing speaker characteristics. Moreover, it ensures consistent model outputs across different runs, effectively mitigating the issue of speaker shifting that may occur between various model executions.
The Perceiver offers the advantage of utilizing multiple references without imposing length restrictions. This flexibility enables it to comprehensively capture diverse aspects of the target speaker. Furthermore, it even allows for the integration of features from other speakers, thereby facilitating the creation of a truly unique voice.

\subsection{Speech Decoder}
The last stage is to convert the SpeechLLM output into waveform. One is to utilize a flow matching\cite{mehta2024matcha} or diffusion-based\cite{anastassiou2024seed} model to transform the speech code generated by the SpeechLLM into an intermediate representation, such as the Mel spectrogram\cite{betker2023better}\cite{anastassiou2024seed}. Then, followed by a vocoder, such as the HifiGAN vocoder, to convert the Mel spectrogram into waveform. 
This method can generate high-quality audio, but it suffers from slow inference and faces complexity in achieving streaming. The second approach is to directly convert the SpeechLLM output, conditioned on speaker embedding, into the final waveform. 
We adopts the second approach, based on the BigVGAN2\cite{lee2022bigvgan} vocoder,
directly reconstructing the audio based on the last hidden state of the SpeechLLM, which is conditioned with speaker embedding. The latent sampling rate is 25Hz. It is interpolated to 100Hz and then input into BigVGAN2. Subsequently, the signal is decoded by BigVGAN2 and finally outputs at a frequency of 24KHz.

\section{Experiments}
\subsection{Dataset}
All training data was collected from the internet, with an initial 120,000 hours of raw audio. After voice separation, speaker segmentation, and filtering using Demucs \cite{defossez2019demucs}, we obtained 34,000 hours of high-quality Chinese-English bilingual data. The dataset includes 25,000 hours of Chinese and 9,000 hours of English audio. We then use ASR (Automatic Speech Recognition) to generate pseudo-labels for the corresponding audio. Finally, we emphasize that punctuation marks are added to the ASR results based on text semantics and speech pauses to create the final training texts. This approach allows users to control pauses flexibly, beyond relying solely on text semantics.

\subsection{Experimental Settings}
\subsubsection{Mixed training of Chinese characters and pinyin}
We randomly select 50\% of the training samples. 
For each sample, we randomly pick 20\% of the Chinese characters. 
If a character is not a polyphonic character, we replace it with its corresponding pinyin. 
The replaced text may include Chinese characters, pinyin, English words, and punctuation marks. 
Then, it is directly tokenized by the BPE tokenizer.

\subsubsection{Speech Codec Training}
In the training of the Speech codec, we only replace Vector Quantization with Finite Scalar Quantization, while keeping other model configurations unchanged. The FSQ levels are $[8, 8, 8, 6, 5]$, the dimension of the VQ codebook is 512, and it contains 8192 codes. Considering that the size and diversity of the training data might affect the utilization rate of the VQ codebook, we also conduct training on a 6,000 hours subset and the entire training dataset respectively.

\subsubsection{Evaluation Settings}
We evaluate indexTTS on four test sets. 
The first two clean test sets are librispeech\cite{panayotov2015librispeech} and Aishell-1\cite{bu2017aishell} test corpus.
The last two sets are composed of 2,000 Chinese samples and 1,000 English samples selected from the CommonVoice\cite{ardila2019common} test dataset. In each set, each speaker has more than two samples.

During the evaluation, for each sample, one other sample from the same speaker corresponding to this sample is randomly selected as the condition prompt. We use Paraformer\cite{gao2022paraformer} ASR to recognize the synthesis results of the Chinese test set, and for the English test set, we use Whisper-large V3\cite{whisper}. This is to evaluate the content consistency.
Regarding speaker similarity (SS), we utilize the ERes2Net\footnote{\url{https://www.modelscope.cn/models/iic/speech_eres2net_sv_zh-cn_16k-common}} model to extract the speaker embeddings from both the prompt and the generated utterances. 
The raw cosine similarity between these embeddings is then regarded as the measure of speaker similarity. 

Additionally, to evaluate the pronunciation correction capability for polyphonic characters, we constructed a challenging Chinese polyphonic character test set comprising 2,500 entries.

\subsection{Experimental Results}

\subsubsection{Controllability of polyphonic characters}
We conducted tests on 2,500 sentences that contain polyphonic characters. The test results are presented in Table~\ref{tab:polyphonic}. Specifically, the inputs of \textbf{A1} are all characters, there are 465 synthesized audio with pronunciation errors of polyphonic characters. This accounts for 18.6\% of the total. Among these audio with pronunciation errors, 437 of them can be accurately corrected by incorporating the correct pinyin as mixed inputs, as shown in \textbf{A2}, accounting for 94\%. The remaining 28 errors, accounting for 1.1\%, that could not be corrected by pinyin might be because errors introduced by the training data have reinforced the SpeechLLM.

\subsubsection{Evaluate The Codec Quantizer}
We compared VQ and FSQ in terms of codebook utilization under varying training data scales(6k and 34k hours) and evaluate on the above four test sets. Results show that with 6k hours training data, VQ has a 55\% low codebook utilization rate. However, when the training data reaches 34k hours, there is little difference between VQ and FSQ, and VQ's utilization rate can also approach 100\%. 50\% of the tokens cover more than 80\% of the total quantity of the tokens that appear in all training data.

\begin{figure}[th]
  \centering
  \includegraphics[scale=0.47]{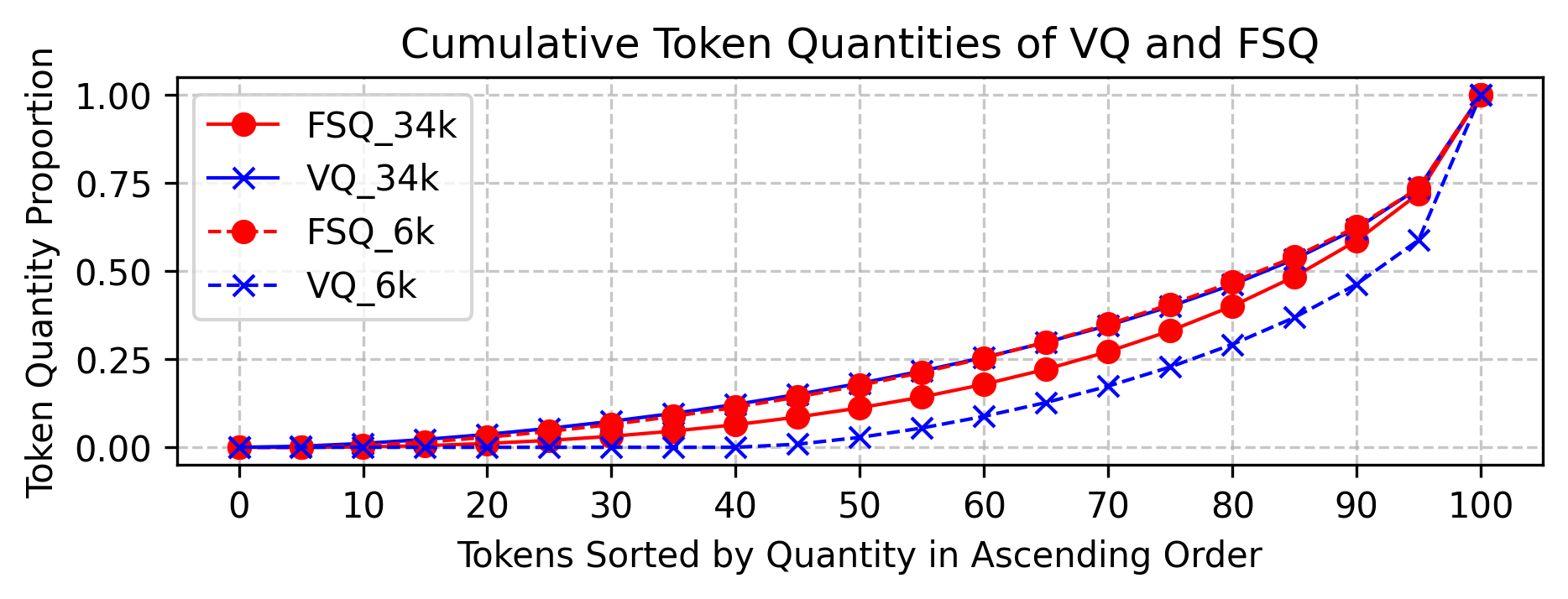}
  \caption{Compare the distribution of codebook utilization rates of VQ and FSQ under different training data scales}
  \label{fig:chart}
\end{figure}

\subsubsection{Comparison Results with Baselines}
We select several top popular zero-shot TTS models in the opensource for comparison, 
including systems XTTS\cite{casanova2024xtts}, CosyVoice2(non-streaming)\cite{du2024cosyvoice2}, FishSpeech\cite{liao2024fish}, FireRedTTS\cite{guo2024fireredtts} and F5-TTS\cite{chen2024f5}. 
The evaluation methodology encompasses both objective and subjective metrics: the word error rate (WER) for the content consistency, the speaker embedding similarity (SS) measure for the evaluation of speech cloning fidelity, and the mean opinion score (MOS) for the quantification of perceptual quality.

The objective evaluation results of WER and SS for IndexTTS and the baseline models across four test sets are presented in Table~\ref{tab:wer_ss_results}. IndexTTS significantly outperforms all other open-source models, demonstrating its robustness and stability. Regarding the SS metric, the performance gap between IndexTTS, CosyVoice2, and F5-TTS is minimal, yet these models exhibit clear advantages over other compared models.

In terms of evaluating the perceptual quality of synthesized audio, we carried out MOS covering three dimensions: prosody, timbre similarity, and sound quality. We conducted a double-blind evaluation by randomly selecting 100 samples from the complete test set to ensure unbiased results. In the subjective evaluation process, we place greater emphasis on the similarity between the synthesized audio and the prompt audio across all aspects. For example, if the sample speeches contain stutters or pauses, we assign a lower score to synthesized results that exhibit overly smooth prosody. Additionally, we also consider the restoration of the sound field characteristics in the prompt audio. The results are shown in Table~\ref{tab:MOS score}, we have outperformed the baseline in nearly all evaluation dimensions, demonstrating significant advantages in timbre similarity and sound quality.

\begin{table}[!htbp]
\caption{MOS Scores for Zero-Shot Cloned Voice}
  \label{tab:MOS score}
  \centering
\begin{tabular}{@{}lcccc@{}}
\toprule
\textbf{Model}       & \multicolumn{1}{c}{\textbf{Prosody}} & \multicolumn{1}{c}{\textbf{Timbre}} & \multicolumn{1}{c}{\textbf{Quality}} & \multicolumn{1}{c}{\textbf{AVG}} \\ \midrule
\textbf{CosyVoice2} & 3.67                                 & 4.05                                   & 3.73                              & 3.81                    \\
\textbf{F5TTS}                & 3.56                                 & 3.88                                   & 3.56                              & 3.66                    \\
\textbf{Fishspeech}           & 3.40                                 & 3.63                                   & 3.69                              & 3.57                    \\
\textbf{FireRedTTS}  & \textbf{3.79}                                 & 3.72                                   & 3.60                              & 3.70                    \\
\textbf{XTTS}                 & 3.23                                 & 2.99                                   & 3.10                              & 3.11                    \\ \cmidrule(){1-5}
\textbf{IndexTTS}        & \textbf{3.79}                                 & \textbf{4.20}                                  & \textbf{4.05}                              & \textbf{4.01}                    \\ \bottomrule
\end{tabular}
\end{table}

Moreover, we randomly selected 200 test samples and calculated the total time taken by all models to synthesize these samples and the GPU resource consumption. The results are presented in Table ~\ref{tab:GPU util score}.

\begin{table}[!htbp]
\caption{GPU Utilization Rate and Test Duration in Experimental Evaluation}
  \label{tab:GPU util score}
  \centering
\begin{tabular}{@{}lcc@{}}
\toprule
\textbf{Model}                   & \textbf{Duration(s)} & \textbf{GPU Util} \\ \midrule
\textbf{CosyVoice2} & 805                  & 48.41\%           \\
\textbf{F5TTS}                   & \textbf{320}         & 42.13\%           \\
\textbf{Fishspeech}              & 756                  & 71.43\%           \\
\textbf{FireRedTTS}              & 732                  & 92.65\%           \\
\textbf{XTTS}                    & 488                  & 87.65\%           \\ \cmidrule(){1-3}
\textbf{IndexTTS}                & 397                  & \textbf{28.47\%}  \\ \bottomrule
\end{tabular}
\end{table}

\subsection{Conclusion}
The IndexTTS system we developed is a GPT-style text-to-speech (TTS) model. It is capable of correcting the pronunciation of Chinese characters using pinyin and controlling pauses at any position through punctuation marks. We enhanced multiple modules of the system, including the improvement of speaker condition feature representation, and the integration of BigVGAN2 to optimize audio quality. Trained on tens of thousands of hours of data, our system achieves state-of-the-art performance, outperforming current popular TTS systems such as XTTS, CosyVoice2, Fish-Speech, and F5-TTS.

\subsection{Limitations}
In this work, several limitations should be acknowledged. Currently, our system does not support instructed voice generation and is limited to Chinese and English, with insufficient capability to replicate rich emotional expressions. In future work, we plan to extend the system to support additional languages, enhance emotion replication through methods such as reinforcement learning, and incorporate the ability to control hyper-realistic paralinguistic expressions, including laughter, hesitation, and surprise, in paralinguistic speech generation.

\bibliographystyle{IEEEtran}
\bibliography{mybib}

\end{document}